\documentclass{PoS}
\usepackage{amsmath}
\usepackage{multirow}
\usepackage{amsfonts}

\title{Long-distance effects in rare and radiative $K$ decays}

\ShortTitle{Long-distance effects in rare and radiative $K$ decays}

\author{\speaker{Christopher Smith}  \thanks{Work supported by the German Bundesministerium f\"{u}r Bildung und Forschung.}\\Institut f\"{u}r Theoretische Teilchenphysik, Karlsruhe Institute of Technology, D-76128 Karlsruhe, Germany\\E-mail: \email{chsmith@particle.uni-karlsruhe.de} }

\abstract{The electroweak structures of the FCNC-induced rare and radiative $K$ decays are presented. While the former decays are sensitive to short-distance physics and offer exceptional probes for new physics, the latter are dominated by long-distance physics. Even so, they should not be set aside since they constitute essential ingredients to control the hadronic uncertainties occurring for the rare decays. This is illustrated by systematically reviewing the phenomenological strategies currently in use to cleanly predict the rare $K$ decay rates in the Standard Model.}

\FullConference{2009 KAON International Conference\\
	June 09 - 12, 2009 \\
	Tsukuba, Japan}

\begin{document}

\section{Introduction}

Flavor changing neutral currents (FCNC) are among our best windows into physics beyond the Standard Model (SM). Indeed, they are absent at tree level and thus entirely generated through quantum loops. This gives them a particular sensitivity to electroweak scale physics. Thus, if New Physics (NP) occurs at a scale not much higher, one can expect significant deviations~\cite{GorbahnHere}.

Rare and radiative $K$ decays permit to probe the quark-level $s\rightarrow d$ FCNC transitions, a fundamental step for reconstructing the flavor sector of a NP model~\cite{CERN}. However, the typical scale of $K$ physics is well within the QCD non-perturbative regime, and the difficult task of dealing with hadronic effects has to be faced. For this, the main tool is Chiral Perturbation Theory (ChPT), the effective theory based on the symmetries of QCD which describes the interactions between the light pseudoscalar mesons. The price to pay for not having solved QCD is a series of phenomenological low-energy constants which have to be fixed from experiment.

Here, our purpose is to review the strategies which are or could be used to resolve the hadronic uncertainties occurring in the rare $K^+\rightarrow\pi^+\nu\bar{\nu}$, $K_L\rightarrow\pi^0\nu\bar{\nu}$, $K_{L}\rightarrow\pi^{0}\ell^{+}\ell^{-}$ and $K_{L}\rightarrow\mu^{+}\mu^{-}$ decays, in particular with the help of radiative $K$ decays. This will be done in Secs.~3, 4, and 5. Before that, in the next section, we describe the electroweak structures driving all these decays.

\section{Short and long-distance effects in rare and radiative $K$ decays}

The $s\rightarrow d$ FCNC transitions depicted in Table~\ref{Table1} contribute to both types of decays, whose final states can actually be identical. What distinguishes them is their sensitivity to short-distance (SD) physics. Rare decays are mostly induced by the $c$ and $t$-quark contributions, and possibly by NP. By contrast, radiative decays are fully dominated by the long-distance (LD) $u$-quark contribution, which makes them rather insensitive to NP. They are always driven by photon penguins (barring pure bremsstrahlung-type processes), hence their name. Note that photons can be real or virtual, and some radiative modes with Dalitz pairs are actually more rare than rare decays.

\paragraph{Standard Model electroweak processes:}

Which of the up-type quarks contribute the most, as indicated in Table~\ref{Table1}, is found by combining the behavior of the FCNC loop as a function of the quark mass~\cite{LimHere,BuchallaBL96} with the scalings of the CKM elements. Specifically, the $Z$ penguin ($W$ boxes are understood) is always SD. Its dominant piece comes from a quadratic breaking of $SU(2)_{L}$, leading to a $m_{u,c,t}^{2}/M_{W}^{2}$ behavior for the loop function. The $\gamma$ penguin does not suppress the light-quark contributions, hence is SD only if CP-violating since then $\operatorname{Im}(V_{us}^{\ast}V_{ud})=0$. Finally, the $\gamma\gamma$ penguin always gives CP-conserving LD contributions; its CP-violating part being strongly suppressed by the additional heavy quark propagator. Even though of higher electroweak order, this process can be competitive thanks to the LD enhancement and large CKM coefficients.

The decays in Table~\ref{Table1} are written in terms of the CP-eigenstates $K_{1}$ and $K_{2}$; the neutral mass eigenstates being $K_{L}\sim K_{2}+\varepsilon K_{1}$ and $K_{S}\sim K_{1}+\varepsilon K_{2}$. Neutral $K$ decays thus proceed either directly, for example $K_{L}\rightarrow K_{2}[\rightarrow X]$, or indirectly, $K_{L}\rightarrow\varepsilon K_{1}[\rightarrow X]$. The indirect contribution is always proportional to the small mixing parameter $\varepsilon_{K}=2.228(11)\times10^{-3}$, but is not always negligible since the direct contribution can be significantly suppressed.

The $K^{+}$ decays are not included in Table~\ref{Table1}, since CP is no longer at play. Instead, $K^{+}$ decays receive contributions from the mechanisms of both columns of Table~\ref{Table1}. For instance, $K^{+}\rightarrow\pi^{+}\nu\bar{\nu}$ has a ``CP-violating'' $Z$-penguin contribution dominated by the $t$ and a ``CP-conserving'' $Z$-penguin contribution with $t$ and $c$ quarks. As can be seen in the table, for all  other $K^{+}$ decays, there is a dominant CP-conserving $u$-quark $\gamma$ or $\gamma\gamma$ penguin, so all these decays are radiative.%

\begin{table}[t] \centering
\begin{tabular}
[t]{ccc}\hline
& CP-violating & CP-conserving\\\hline
\centering \includegraphics[width=2cm]{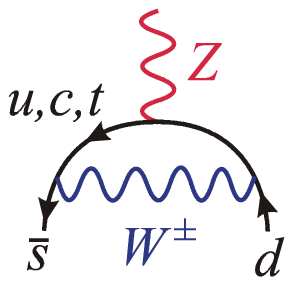}  & $\centering
\begin{array}
[b]{c}%
t\text{-quark dominates:}\\
K_{2}\rightarrow\pi^{0}\nu\bar{\nu}\\
K_{2}\rightarrow\pi^{0}\ell^{+}\ell^{-}\\
K_{1}\rightarrow\ell^{+}\ell^{-}%
\end{array}
$ & $\centering
\begin{array}
[b]{c}%
t\text{, }c\text{ (+ small }u\text{) quarks:}\\
K_{1}\rightarrow\pi^{0}\nu\bar{\nu}\\
K_{1}\rightarrow\pi^{0}\ell^{+}\ell^{-}\\
K_{2}\rightarrow\ell^{+}\ell^{-}%
\end{array}
$\\\hline
\centering \includegraphics[width=2cm]{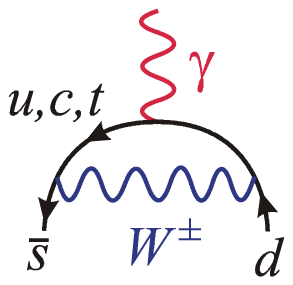}  & $\centering
\begin{array}
[b]{c}%
\text{Only }t\text{ \& }c\text{ quarks:}\\
K_{2}\rightarrow\pi^{0}\ell^{+}\ell^{-}\\
K\rightarrow\pi\pi\gamma\\
\;
\end{array}
$ & $\centering
\begin{array}
[b]{c}%
u\text{ quark dominates:}\\
K_{1}\rightarrow\pi^{0}\ell^{+}\ell^{-}\\
K\rightarrow\pi\pi\gamma\\
\;
\end{array}
$\\\hline
\centering \includegraphics[width=2cm]{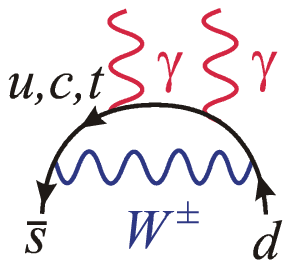}  & $\centering
\begin{array}
[b]{c}%
\text{Only }t\text{ \& }c\text{ quarks:}\\
\text{negligible}\\
\text{ (}\sim1/m_{c,t}\text{)}\\
\;
\end{array}
$ & $\centering
\begin{array}
[b]{c}%
u\text{ quark dominates:}\\
K_{1,2}\rightarrow\gamma\gamma,\;K_{1,2}\rightarrow\pi^{0}\gamma\gamma\\
K_{1,2}\rightarrow\pi^{0}\ell^{+}\ell^{-}\\
K_{1,2}\rightarrow\ell^{+}\ell^{-}%
\end{array}
$\\\hline
\end{tabular}
$\ $%
\caption{Electroweak processes relevant for rare and radiative $K$ decays, with the dominant contributions for each mode ($\gamma$ in the final state stands for both a real photon or an $\ell^+\ell^-$ Dalitz pair, and $\ell = e, \mu$).}
\label{Table1}
\end{table}

\paragraph{Windows for New Physics:}

In summary, the only modes giving us access to SD physics are: $K^{+}\rightarrow\pi^{+}\nu\bar{\nu}$ and $K_{L}\rightarrow\pi^{0}\nu\bar{\nu}$ ($K_{S}\rightarrow\pi^{0}\nu\bar{\nu}$ is difficult to measure), induced by the $Z$ penguin, $K_{L}\rightarrow\pi^{0}e^{+}e^{-}$ and $K_{L}\rightarrow\pi^{0}\mu^{+}\mu^{-}$, which receive contributions from the three electroweak processes, and $K_{L}\rightarrow\mu^{+}\mu^{-}$, for which the $\gamma$ penguin is absent (the helicity suppression makes $K_{L}\rightarrow e^{+}e^{-}$ very small). These modes are naturally the most sensitive to NP effects (see e.g.~\cite{GorbahnHere,CERN,NP,MesciaST06} for the impacts of supersymmetry). Further, taken in combination, they offer a powerful discriminating tool thanks to their different sensitivities to the underlying electroweak processes.

In principle, the SD parts can also be accessed through CP asymmetries. As will be briefly discussed in Sec.~4, this is especially clean for the $K^{+}$ modes, for which the direct CP-asymmetries typically arise through the interferences between the ``CP-violating'' and ``CP-conserving'' pieces of Table 1. For the neutral modes, CP asymmetries are often driven by the mixing and essentially proportional to $\varepsilon_{K}$, with small corrections from direct CP-violating effects.

\section{Long-distance effects in $K^{+}\rightarrow\pi^{+}\nu\bar{\nu}$ and $K_{L}\rightarrow\pi^{0}\nu\bar{\nu}$}

\paragraph{Matrix elements:}

\begin{figure}[ptb]
\centering    \includegraphics[width=0.94\textwidth]{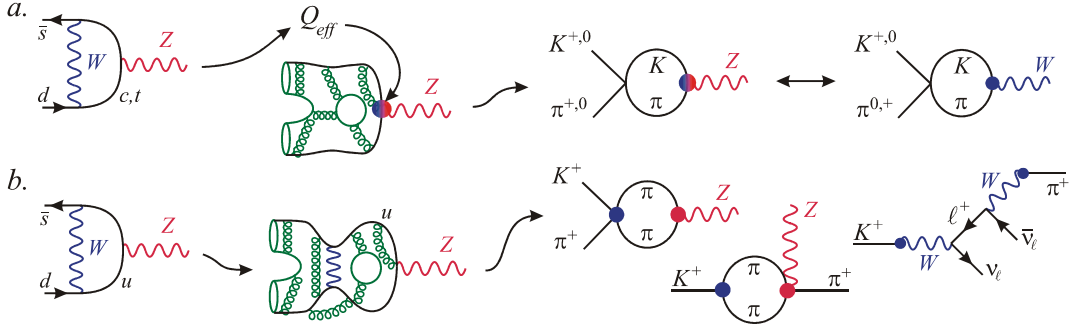} 
\caption{LD effects in $K\rightarrow\pi\nu\bar{\nu}$. $a$: The matrix elements of $Q_{eff}$ and their relation to those of the charged current. $b$: The up-quark $Z$ penguin and associated LD processes.}%
\label{Fig1}%
\end{figure}

Even if the $t$ and $c$-quark $Z$ penguins are said to be SD, there are hadronic effects to be dealt with. Indeed, these SD contributions are encoded as effective four-fermion operators, of which we need the matrix elements between hadron states to get down to physical observables (see Fig.~\ref{Fig1}$a$). In the SM, there is only one dimension-six operator~\cite{GorbahnHere}:
\begin{equation}
Q_{eff}=\bar{s}\gamma^{\mu}\left(  1-\gamma_{5}\right)  d\otimes\bar{\nu}\gamma_{\mu}\left(  1-\gamma_{5}\right)  \nu\;.
\end{equation}
To obtain $\langle\pi\nu\bar{\nu}|Q_{eff}|K\rangle$, we need the matrix element of the vector current, $\langle\pi|\bar{s}\gamma^{\mu}d|K\rangle$, which is described in terms of the vector and scalar form-factors (only the former is needed for massless neutrinos). These form-factors are directly related to those of the charged current decays $K\rightarrow\pi\ell\nu$ (so-called $K_{\ell3}$), induced by the Fermi interaction:%
\begin{equation}
Q_{Fermi}=\bar{s}\gamma^{\mu}\left(  1-\gamma_{5}\right)  u\otimes\bar{\nu}\gamma_{\mu}\left(  1-\gamma_{5}\right)  \ell\;.\label{Fermi}%
\end{equation}
In the isospin limit, $\langle\pi|\bar{s}\gamma^{\mu}u|K\rangle=\langle\pi|\bar{s}\gamma^{\mu}d|K\rangle$, up to simple Clebsch-Gordan coefficients.

Isospin-breaking effects have to be accounted for given the percent level of precision we are after. For that, three very clean theoretical ratios are used, involving the vector form-factors $f_{+}^{ij}(q^{2})$ of the $\langle \pi^{i}|\bar{s}\gamma^{\mu}u|K^{j}\rangle$ and $\langle\pi^{i}|\bar{s}\gamma^{\mu}d|K^{j}\rangle$ matrix elements ($q^{2}$ is the momentum transfer), as well as their slopes $\lambda_{+}^{ij}$ defined as the coefficients of their Taylor expansions around $q^{2}=0$~\cite{MesciaS07,BijnensG07}:%
\begin{equation}
r=\frac{f_{+}^{+0}(q^{2})f_{+}^{00}(q^{2})}{f_{+}^{++}(q^{2})f_{+}^{0+}(q^{2})}=1.0000(2),\;\;r_{K}=\frac{f_{+}^{++}(0)}{f_{+}^{0+}(0)}=1.0015(7),\;\;r_{\lambda}=\frac{\lambda_{+}^{++,00}}{\lambda_{+}^{+0,0+}}=0.990(5)\;.
\end{equation}
Using the results on the form-factors of the Flavianet fit to $K_{\ell3}$ data~\cite{FlaviaNet}, the phase-space integrated matrix elements $\kappa_{\nu}\sim\int d\Phi_{3}|\langle\pi\nu\bar{\nu}|Q_{eff}|K\rangle|^{2}$  (see~\cite{MesciaS07} for the precise definition) are obtained with an impressive precision: $\kappa_{\nu}^{+}=0.5168\left(25\right)  \times10^{-10}$ and $\kappa_{\nu}^{0}=2.190\left(  18\right)\times10^{-10}$, with the errors coming mostly from $K_{\ell3}$ data. Consequently, the errors from the LD effects on the $t$ and $c$-quark contributions to $K\rightarrow \pi\nu\bar{\nu}$ amount to only a few percents of the total error~\cite{GorbahnHere}.

\paragraph{Up-quark $Z$ penguin:}

The $u$-quark contribution to the $Z$ penguin is a priori suppressed compared to that of the $c$-quark by the tiny $m_{u}^{2}/m_{c}^{2}$ ratio. However, since the $u$-quark is dynamical, this translates as a significant $\mathcal{O}(m_{K}^{2})/m_{c}^{2}\sim10\%$ correction. At this level, the other LD contributions arising from the $W$ box or to the $\Delta S=1\otimes\Delta S=1$ tree-level processes also have to be included~\cite{IsidoriMS05,KamenikS09}, see Fig.~\ref{Fig1}$b$. All these LD contributions are CP-conserving, hence do not occur for $K_{L}\rightarrow\pi^{0}\nu\bar{\nu}$.

The computation is done in ChPT by including the hadronized $\Delta S=1$ three-flavor effective operators $Q_{1}$ to $Q_{6}$ (in terms of the weak low-energy constants $G_{8}$ and $G_{27}$, fixed from $K\rightarrow\pi\pi$ and thus accounting for the $\Delta I=1/2$ enhancement), together with the hadronized effective interactions $\bar{u}d,\bar{u}s\rightarrow W\rightarrow \bar{\ell}\nu$ and $\bar{u}u,\bar{d}d,\bar{s}s\rightarrow Z\rightarrow\bar {\nu}\nu$. A complication is that the standard prescription to include the neutral current within ChPT by promoting ordinary derivatives to covariant ones does not work in presence of $\Delta S=1$ interactions, because the GIM mechanism is missed. As a result, the contributions from $\langle\pi\nu\bar{\nu}|Q_{eff}|K\rangle$ described in the previous section are entangled with the genuine $u$-quark $Z$ penguin effects (in other words, ChPT regenerates the meson loop of Fig.~\ref{Fig1}$a$).

A simple procedure to enforce the GIM mechanism is to require the absence of a local $K_{L}\rightarrow Z$ coupling~\cite{IsidoriMS05}, so that the meson loops have the expected $u$-quark $Z$ penguin electroweak structure. Many counterterms occur, some of them related to $K^{+}\rightarrow\pi^{+}\gamma^{\ast}[\rightarrow\ell^{+}\ell^{-}]$. Allowing for a conservative uncertainty for those which are unknown, the correction from the light degrees of freedom is $\delta P_{u}=0.04(2)$~\cite{IsidoriMS05}, to be compared to the $c$-quark contribution $P_{c}=0.372(15)$~\cite{GorbahnHere,RareK}.

At this point, the other correction of order $m_{K}^{2}/m_{c}^{2}$ can be evaluated. It arises from dimension eight operators induced by the $c$-quark, i.e. four-fermion operators with two derivatives (each counting as $\mathcal{O}(m_{K})$ given the momentum scale involved). The problem reduces to the evaluation of the matrix elements of these operators. Though this cannot be done exactly, an approximate matching with the non-local $u$-quark $Z$ penguin computed in ChPT is possible, making use of the CKM unitarity (the $t$-quark piece is negligible). It turns out that this correction $\delta P_{c}$ is smaller than $\delta P_{u}$ as it misses the $\Delta I = 1/2$ enhancement. The combined correction is then simply $\delta P_{u,c}=0.04(2)$ and amount to a $6\%$ increase of the $K^{+}\rightarrow\pi^{+}\nu\bar{\nu}$ rate~\cite{IsidoriMS05}.

\section{Long-distance effects in $K_{L}\rightarrow\pi^{0}e^{+}e^{-}$ and $K_{L}\rightarrow\pi^{0}\mu^{+}\mu^{-}$}

These decays receive three contributions (see e.g.~\cite{MesciaST06}). The first piece is the so-called direct CP-violation (DCPV), induced by the $Z$ and $\gamma$ penguins with $t$ and $c$ quarks (see Table~\ref{Table1}). The second piece is the indirect CP-violation (ICPV), which proceeds through the small $K_{1}$ component of the $K_{L}$. The $K_{1}\rightarrow\pi^{0}\ell^{+}\ell^{-}$ decay is CP-conserving and dominated by the LD $u$-quark $\gamma$ penguin. Both the DCPV and ICPV can produce the lepton pair in a $1^{--}$ state, and thus interfere. Finally, the third piece conserves CP (CPC) , and proceeds through a purely LD $\gamma\gamma$ penguin.

Each of these three pieces involves LD physics at some stage. Let us describe the strategies used to bring them under control.

\begin{figure}[t]
\centering  \includegraphics[width=0.95\textwidth]{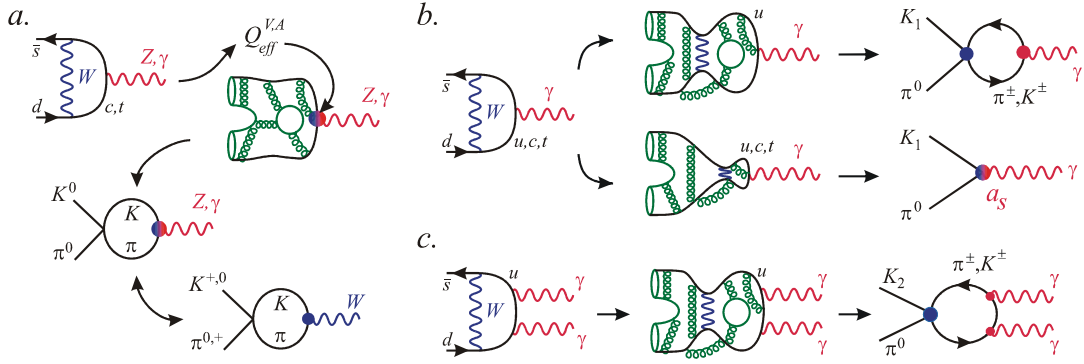} \caption{LD effects in $K_{L}\rightarrow\pi^{0}\ell^{+}\ell^{-}$. $a$: The matrix elements of $Q_{eff}^{V,A}$ and their relation to those of the charged current (DCPV). $b$ and $c$: Meson representations for the $\gamma$ and $\gamma\gamma$ penguins (ICPV and CPC, resp.).}%
\label{Fig2}%
\end{figure}

\paragraph{Matrix elements for DCPV:}

The situation is similar to that for $K\rightarrow\pi\nu\bar{\nu}$, with the SD $Z$ and $\gamma$ penguins with $t$ and $c$ quarks encoded into two effective dimension-six operators~\cite{BuchallaBL96},%
\begin{equation}
Q_{eff}^{V}=\bar{s}\gamma^{\mu}\left(  1-\gamma_{5}\right)  d\otimes\bar{\ell}\gamma_{\mu}\ell\;,\;\;Q_{eff}^{A}=\bar{s}\gamma^{\mu}\left(  1-\gamma_{5}\right)  d\otimes\bar{\ell}\gamma_{\mu}\gamma_{5}\ell\;.\label{VA}%
\end{equation}
The hadronic matrix elements are the same as for $K\rightarrow\pi\nu\bar{\nu}$, though now the scalar form-factor contributes to $\langle\pi^{0}\mu^{+}\mu^{-}|Q^A_{eff}|K_{L}\rangle$ (its impact for $\pi^{0}e^{+}e^{-}$ is suppressed by the electron mass). Isospin-breaking effects can be controlled as before (Fig.~\ref{Fig2}$a$), since the scalar $f_{0}^{ij}(q^{2})$ and vector $f_{+}^{ij}(q^{2})$ form-factors are equal at $q^{2}=0$, while the scalar slope is not very precisely known but has a small impact. Using the Flavianet fit~\cite{FlaviaNet}, the phase-space integrated matrix elements $\kappa_{\ell}^{V,A}\sim\int d\Phi_{3}|\langle\pi^{0}\ell^{+}\ell^{-}|Q_{eff}^{V,A}|K_{L}\rangle|^{2}$ are found with a few per mil precision (see~\cite{MesciaS07} for details). This is more than enough given the larger errors on the other two contributions to $K_{L}\rightarrow\pi^{0}\ell^{+}\ell^{-}$.

\paragraph{ICPV contribution:}

The process $K_{L}\rightarrow\varepsilon K_{1}[\rightarrow\pi^{0}\ell^{+}\ell^{-}]$ is related to $K_{S}\approx K_{1}\rightarrow\pi^{0}\ell^{+}\ell^{-}$, which is dominated by LD physics (Fig.~\ref{Fig2}$b$). The ChPT analysis (supplemented with dispersion relations) of Ref.~\cite{DambrosioEIP98} has shown that meson loops are subleading and the amplitude is essentially constant, proportional to a linear combination of counterterms denoted $a_{S}$. Using the experimental $K_{S}\rightarrow\pi^{0}e^{+}e^{-}$ and $K_{S}\rightarrow\pi^{0}\mu^{+}\mu^{-}$ rates~\cite{NA48_KSpill} gives $|a_{S}|=1.2(2)$. The sign of $a_{S}$ is inaccessible since loops are small and these rates are simply proportional to $a_{S}^2$.

This 20\% uncertainty on $a_{S}$ is the main source of error for the $K_{L}\rightarrow\pi^{0}\ell^{+}\ell^{-}$ rates~\cite{MesciaST06}. Further, the sign of $a_{S}$ determines that of the ICPV and DCPV interference, thus unknown at present. To improve our knowledge of $a_{S}$, there are two possible approaches besides better measurements of $K_{S}\rightarrow\pi^{0}\ell^{+}\ell^{-}$. First, $a_{S}$ can be related to the counterterm $a_{+}$ occurring for $K^{+}\rightarrow\pi^{+}\ell^{+}\ell^{-}$, though in a somewhat model-dependent way~\cite{BuchallaDI03,FGD04}. Second, the decay $K_{L}\rightarrow\pi^{0}\pi^{0}\ell^{+}\ell^{-}$ depends on the same counterterm $a_{S}$, and has a non-negligible loop contribution~\cite{FunckK93}. Though its branching ratio is small, of $\mathcal{O}(10^{-9})$ for $\ell=e$, its measurement would give us both the size and sign of $a_{S}$.

Finally, it should be mentioned that, if statistics is sufficient, the rare decay $K_{L}\rightarrow\pi^{0}\mu^{+}\mu^{-}$ itself could fix the sign of $a_{S}$ through the observation of a forward-backward asymmetry~\cite{MesciaST06}, which is predicted to be of about $\operatorname{sign}(a_{S})\times15\%$.

\paragraph{CPC contribution:}

The final contribution is induced by the $\gamma\gamma$ penguin (Fig.~\ref{Fig2}$c$), and is purely LD. In ChPT, it is represented by $\pi^{\pm}$ and $K^{\pm}$ loops, is finite at $\mathcal{O}(p^{4})$, and produces the lepton pair in a $0^{++}$ state only. It is helicity suppressed and thus relevant only for $K_{L}\rightarrow\pi^{0}\mu^{+}\mu^{-}$.

To get a handle on higher order effects, the experimental information on the $K_{L}\rightarrow\pi^{0}\gamma\gamma$ mode can be used (as well as on $K_{S}\rightarrow\gamma\gamma$, which involves the same $\mathcal{O}(p^{6})$ counterterms). For the muon mode, the $K_{L}\rightarrow\pi^{0}\gamma\gamma$ rate permits to partially account for $\mathcal{O}(p^{6})$ effects, bringing the CPC piece under control to within 30\%~\cite{IsidoriSU04}. For the electron mode, the absence of the $\gamma\gamma$ contribution is confirmed by the photon energy spectrum in $K_{L}\rightarrow\pi^{0}\gamma\gamma$. Indeed, to escape the helicity suppression, the $e^{+}e^{-}$ pair is necessarily in a $2^{++}$ tensor state. This requires a significant production of the two photons also in a $2^{++}$ state, which would have had a clear signature on the spectrum~\cite{BuchallaDI03}.

\paragraph{Direct CP-asymmetries:}

Radiative $K^{+}$ decays are LD-dominated, yet the asymmetries%
\begin{equation}
A_{CP}\left(  K^{+}\rightarrow X^{+}\right)  =\frac{\Gamma\left(K^{+}\rightarrow X^{+}\right)  -\Gamma\left(  K^{-}\rightarrow X^{-}\right)}{\Gamma\left(  K^{+}\rightarrow X^{+}\right)  +\Gamma\left(  K^{-}\rightarrow X^{-}\right)  }\;
\end{equation}
are sensitive to SD physics. Possible channels are for example $X^{+}=\pi^{+}\ell^{+}\ell^{-}$~\cite{DambrosioEIP98} or $X^{+}=\pi^{+}\pi^{0}\gamma$~\cite{DAmbrosioI96}. However, these asymmetries are significantly suppressed in the SM by the smallness of the SD parts relative to the LD contributions, and should not exceed about $10^{-4}$. Phase-space asymmetries, e.g. the $K_{L,S}\rightarrow\pi^{0}\mu^{+}\mu^{-}$ forward-backward asymmetry or the $K_{L}\rightarrow\pi^{+}\pi^{-}\gamma^{\ast}$ angular asymmetry, could also be interesting even though one needs to disentangle indirect and direct CP violation~\cite{MesciaST06,DambrosioEIP98}.

\section{Long-distance effects in $K_{L}\rightarrow\mu^{+}\mu^{-}$}

\begin{figure}[t]
\centering    \includegraphics[width=0.94\textwidth]{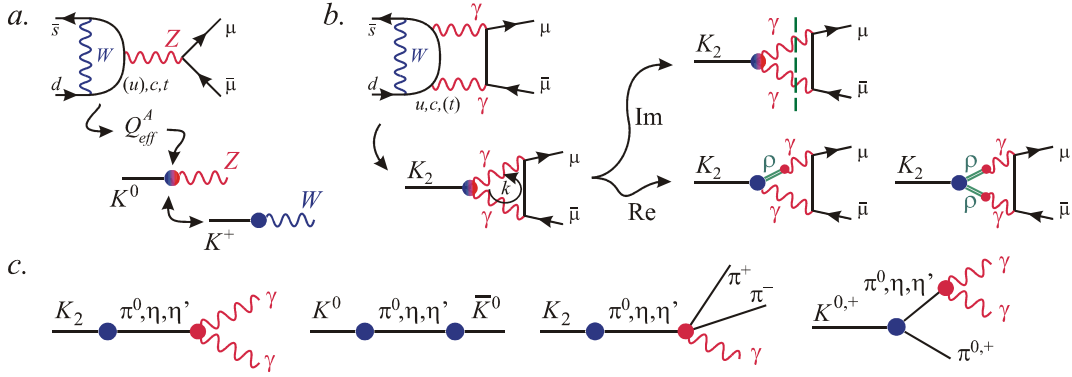} \caption{LD effects in $K_{L}\rightarrow\mu^{+}\mu^{-}$. $a$: The matrix element of $Q_{eff}^{A}$ and its relation to that of the charged current. $b$: Meson representation for the $\gamma\gamma$ penguin. $c$: The pole amplitudes sensitive to $G_{8}^{s}$.}%
\label{Fig3}%
\end{figure}

At first sight, the absence of the $\gamma$ penguin appears to promise a relatively simple structure, and thereby a good control on the hadronic uncertainties. Indeed, the $Z$ penguin is as always dominated by SD, and the decay being CP-conserving, it receives both $t$ and $c$-quark contributions (see Table~\ref{Table1} and Fig.~\ref{Fig3}$a$). However, this picture is upset by the large contribution from the $\gamma\gamma$ penguin (Fig.~\ref{Fig3}$b$), which gives rise to three problems to be addressed before being able to use the measurement $B(K_{L}\rightarrow\mu^{+}\mu^{-})=6.87(11)\cdot10^{-9}$ to constrain SD physics.

First, the absorptive part of the $\gamma\gamma$ loop, precisely known from $K_{L}\rightarrow\gamma\gamma$, contributes for $6.7\cdot10^{-9}$ and nearly saturates the branching ratio. This clearly makes the current experimental precision on $B(K_{L}\rightarrow\mu^{+}\mu^{-})$ insufficient to access to the interesting SD physics. Second, the $\gamma\gamma$ loop is divergent in ChPT, in stark contrast to $K_{L}\rightarrow\pi^{0}\ell^{+}\ell^{-}$ where it is finite at leading order. This means that unknown counterterms occur and the dispersive part of the $\gamma\gamma$ loop cannot be predicted within ChPT. Finally, this dispersive part interferes with the SD piece, but the sign is unknown.

Though a definitive answer is still lacking, let us detail the current and possible future strategies designed to bring this mode under control.

\paragraph{Matrix element:}

The effective operators are the same as for $K_{L}\rightarrow\pi^{0}\ell^{+}\ell^{-}$, but with different SD coefficients since the $\gamma$ penguin is absent~\cite{GorbahnH06}. Only the $\langle\mu^{+}\mu^{-}|Q_{eff}^{A}|K_{L}\rangle$ matrix element is non-zero, and is again related to that of the charged current $\langle\mu^{+}\nu_{\mu}|Q_{Fermi}|K^{+}\rangle$ which, by PCAC, is given in terms of the $K^{+}$ decay constant (Fig.~\ref{Fig3}$a$). Isospin-breaking effects could be included but are not currently needed given the uncertainties on the other contributions.

\paragraph{Sign of the $\gamma\gamma$ contribution:}

To get this sign, we need that of the $K_{L}\rightarrow\gamma\gamma$ amplitude. However, its evaluation is rather tricky because it vanishes at leading order in ChPT. This cancellation occurs because the operator driving $K_{L}\rightarrow\gamma\gamma$ is $Q_{1}=\bar{s}\gamma^{\mu}(1-\gamma_{5})d\otimes\bar{u}\gamma_{\mu}(1-\gamma_{5})u$, and in a pole model (Fig.~\ref{Fig3}$c$), there is no way to get an $\bar{u}u$ state as a linear combination of $\pi^{0}$ and $\eta_{8}$ states~\cite{GerardST05}. The same mechanism is at play for $\Delta M_{LS}$ or for the non-local magnetic contribution to $K_{L}\rightarrow\pi^{+}\pi^{-}\gamma$.

To circumvent this problem and consistently account for the class of higher order corrections induced by the singlet $\eta_{1}$ in ChPT, the best strategy is to first go to $U(3)$ ChPT, where the pole amplitudes no longer vanish, and then project back onto standard $SU(3)$ ChPT. Doing so for $K_{L}\rightarrow\gamma\gamma$, the amplitude turns out to be proportional to a new weak low-energy constant $G_{8}^{s}$, of which we now need the sign. To this end, one can either make some generic assumptions on the behavior of the $Q_{1}$ to $Q_{6}$ operators in the non-perturbative regime, or constrain $G_{8}^{s}$ experimentally from those pole amplitudes which do not vanish at leading order
in $SU(3)$ ChPT, for example $K_{S}\rightarrow\pi^{0}\gamma\gamma$ or $K^{+}\rightarrow\pi^{+}\gamma\gamma$ (Fig.~\ref{Fig3}$c$), see Ref.~\cite{GerardST05} for more details.

\paragraph{Size of the $\gamma\gamma$ contribution:}

The problem reduces to that of computing the value of a low-energy constant. This is very difficult as one must deal with hadronic physics above the ChPT scale but still well within the non-perturbative regime. The standard approach is to parametrize the form-factor for the vertex $K_{L}\rightarrow\gamma^{\ast}\gamma^{\ast}$, and then use this form-factor to compute the $\gamma\gamma$ loop. For moderate photon virtuality, one relies on Vector Meson Dominance (VMD) or large $N_{C}$ arguments (Fig.~\ref{Fig3}$b$), while at high virtuality, an approximate matching with the partonic $u$ and $c$-quark $\gamma\gamma$ penguin is performed (see e.g.~\cite{IsidoriU03} and references there).

The VMD ansatz introduces a number of free parameters which have to be fixed from the $K_{L}\rightarrow\gamma e^{+}e^{-}$, $K_{L}\rightarrow\gamma\mu^{+}\mu^{-}$ and $K_{L}\rightarrow e^{+}e^{-}\mu^{+}\mu^{-}$ differential rates. This can only be done partially at present, and improving the experimental accuracy would be welcome, especially on the latter mode. Kinematically, $K_{L}\rightarrow\mu^{+}\mu^{-}\mu^{+}\mu^{-}$ would be even better but its $10^{-13}$ branching ratio~\cite{ZhangG98} presumably forbids any study of the differential rate.

Altogether, the dispersive $\gamma\gamma$ contribution is similar in size to the SD part, but is poorly known and thus the main source of uncertainty on the $K_L\rightarrow \mu^+\mu^-$ rate in the SM~\cite{IsidoriU03}.%

\begin{table}[t] \centering
\begin{tabular}
[t]{lll}\hline
\;NP probe\;\;\; & \ \ Rare decays & Decays to be used to control LD effects\\\hline
\multirow{4}{*}{\parbox[l]{1em}%
{\includegraphics[trim = 6mm 0mm 0mm 0mm, clip, width=2.3cm]{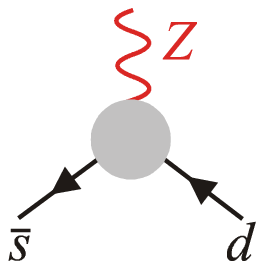}}}  &
\ \ $K\rightarrow\pi\nu\bar{\nu}$ & $K_{\ell3}$\; (matrix elements)\\
&  & $K^{+}\rightarrow\pi^{+}\ell^{+}\ell^{-}$\; (up-quark $Z$ penguin)\\
&  & \\
&  & \\\hline
\multirow{4}{*}{\parbox[l]{1em}%
{\includegraphics[trim = 6mm 0mm 0mm 0mm, clip, width=2.3cm]{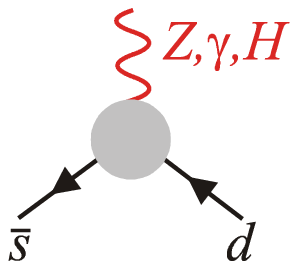}}}  &
\ \ $K_{L}\rightarrow\pi^{0}\ell^{+}\ell^{-}$ & $K_{\ell3}$\; (matrix
elements, DCPV)\\
&  & $K_{S}\rightarrow\pi^{0}\ell^{+}\ell^{-},\;K^{+}\rightarrow\pi^{+}\ell
^{+}\ell^{-},\;K_{L}\rightarrow\pi^{0}\pi^{0}\ell^{+}\ell^{-}$\; (ICPV)\\
&  & $K_{L}\rightarrow\pi^{0}\gamma\gamma,\;K_{S}\rightarrow\gamma\gamma$\;
(CPC)\\
& \multicolumn{2}{l}{\;\;\;(also accessible with $K^{+}\rightarrow\pi^{+}\ell^{+}%
\ell^{-},\;K^{+}\rightarrow\pi^{+}\pi^{0}\gamma,...$ CP-asymmetries)}\\\hline
\multirow{4}{*}{\parbox[l]{1em}%
{\includegraphics[trim = 6mm 0mm 0mm 0mm, clip, width=2.3cm]{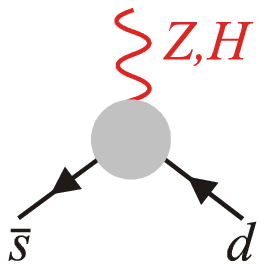}}}  &
\ \ $K_{L}\rightarrow\mu^{+}\mu^{-}$ & $K_{\ell2}$\; (matrix elements)\\
&  & $K_{L}\rightarrow\gamma\gamma,\;K_{S}\rightarrow\pi^{0}\gamma\gamma
,\;K^{+}\rightarrow\pi^{+}\gamma\gamma$\; (interference sign)\\
&  & $K_{L}\rightarrow\gamma\ell^{+}\ell^{-},\;K_{L}\rightarrow\ell^{+}\ell
^{-}\ell^{\prime+}\ell^{\prime-}$\; (dispersive $\gamma\gamma$)\\
&  & \\\hline
\end{tabular}
$\ $%
\caption{Sensitivities of the rare $K$ decays to NP entering $Z$, $\gamma$ and non-standard Higgs penguins~\cite{GorbahnHere,CERN,NP,MesciaST06}. The last column lists the modes used to control their hadronic uncertainties.}
\label{Table2}
\end{table}

\section{Conclusion}

The rare decays $K\rightarrow\pi\nu\bar{\nu}$, $K_{L}\rightarrow\pi^{0}\ell^{+}\ell^{-}$, and to some extent $K_{L}\rightarrow\mu^{+}\mu^{-}$ offer unique windows into $s\rightarrow d$ transitions thanks to their sensitivity to SD physics, and thereby to possible NP effects. In addition, each mode can probe specific electroweak processes, and thus, when taken in combination, these decays form a powerful discriminating tool (see Table~\ref{Table2}). Even if NP is first discovered at the LHC, they will remain essential to uncover its flavor structures.

In this program, radiative decays have an important supporting role. Being dominated by LD physics, they are mostly insensitive to NP. This makes them appropriate to reliably fix the hadronic quantities originating from residual LD contributions in rare $K$ decays (see Table~\ref{Table2}). The exceptional cleanness of the rare modes would not be possible without experimental information on the radiative ones (as well as on the $K_{\ell3}$ decays). Therefore, besides the golden rare decay modes, radiative decays should be an integral part of the experimental programs under development~\cite{CKM}, and will ensure a rich and fruitful $K$ physics phenomenology for the years to come.

\end{document}